\documentclass[12pt]{iopart}

\usepackage{graphicx}

\begin{document}

\def\ms {\overline{\rm MS}}
\def\dis{{\rm DIS}_{\gamma}}
\def\d{{\rm d}}

\def\lp{\left. }
\def\rp{\right. }
\def\lr{\left( }
\def\rr{\right) }
\def\le{\left[ }
\def\re{\right] }
\def\lg{\left\{ }
\def\rg{\right\} }
\def\lb{\left| }
\def\rb{\right| }

\def\beq{\begin{equation}}
\def\eeq{\end{equation}}
\def\bea{\begin{eqnarray}}
\def\eea{\end{eqnarray}}

\title[Hard Photoproduction at HERA]{Hard Photoproduction at HERA}

\author{M Klasen}

\address{Laboratoire de Physique Subatomique et de Cosmologie, Universit\'e
 Joseph Fourier/CNRS-IN2P3, 53 Avenue des Martyrs, F-38026 Grenoble, France}
\ead{klasen@lpsc.in2p3.fr}
\begin{abstract}
In view of possible photoproduction studies in ultraperipheral heavy-ion
collisions at the LHC, we briefly review the present theoretical
understanding of photons and hard photoproduction processes at HERA,
discussing the production of jets, light and heavy hadrons, quarkonia, and
prompt photons. We address in particular the extraction of the strong
coupling constant from photon structure function and inclusive jet
measurements, the infrared safety and computing time of jet definitions, the
sensitivity of dijet cross sections on the parton densities in the photon,
factorization breaking in diffractive dijet production, the treatment of the
heavy-quark mass in charm production, the relevance of the color-octet
mechanism for quarkonium production, and isolation criteria for prompt
photons.
\end{abstract}

\vspace*{-125mm}
\begin{flushright}
LPSC 07-006
\end{flushright}

\maketitle


\section{Introduction}

\begin{figure}
 \centering
 \includegraphics[width=.66\columnwidth]{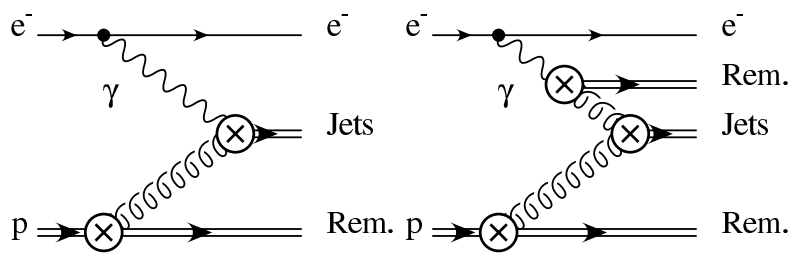}
 \caption{\label{fig:1}Factorization of direct (left) and resolved (right)
 photoproduction in the QCD-improved parton model.}
\end{figure}

Electron-proton scattering at HERA is dominated by the exchange of
low-virtuality (almost real) photons \cite{kla02}. If the electron is
anti-tagged or tagged at small angles, the photon flux from the electron can
be calculated in the Weizs\"acker-Williams approximation, where the energy
spectrum of the exchanged photons is given by
\bea
 \hspace*{-10mm}
 f_{\gamma/e}^{\rm brems}(x)&=&\frac{\alpha}{2\pi}\left[
 \frac{1+(1-x)^2}{x}\ln\frac{Q^2_{\max}(1-x)}{m_e^2 x^2}
 + 2 m_e^2 x\left(\frac{1}{Q^2_{\max}}-\frac{1-x}{m_e^2 x^2}\right)
 \right]\hspace*{10mm}
\eea
and the subleading non-logarithmic terms modify the cross section typically
by 5\% \cite{kes75}. In the QCD-improved parton model, valid for hard
scatterings, the photons can then interact either directly with the partons
in the proton (Fig.\ \ref{fig:1}, left) or resolve into a hadronic
structure, so that
%
%
their own partonic constituents interact with the partons in the proton
(Fig.\ \ref{fig:1}, right). While this separation is valid at leading order
(LO) in QCD perturbation theory, the two processes are intimately linked at
next-to-leading order (NLO) through the mandatory factorization of a
collinear singularity, that arises from the splitting of the photon into a
quark-antiquark pair and induces a mutual logarithmic factorization scale
dependence in both processes. In close analogy to deep-inelastic
electron-proton scattering, one can define a photon structure function
\bea
 \hspace*{-25mm}
 F_2^{\gamma}(Q^2)&=&\sum_q 2 x e_q^2\lg f_{q/\gamma}(Q^2)+\frac{\alpha_s(Q^2)}
 {2\pi}\le C_{q}\otimes f_{q/\gamma}(Q^2)
 + C_{g}\otimes f_{g/\gamma}(Q^2)\re
 +\frac{\alpha}{2\pi}e_q^2C_{\gamma}\rg
\eea
that is related to the parton densities in the photon and has been measured
in electron-positron collisions at LEP. Even the strong coupling constant
$\alpha_s$ that appears in the expression above can be determined rather
precisely in fits to these data \cite{alb02}. A convenient modification of
the $\ms$ factorization scheme consists in absorbing the point-like Wilson
coefficient
\bea
 \hspace*{-10mm}
 C_{\gamma}(x)&=&2N_C\, C_g(x)=3\, \left[ \left( x^2+(1-x)^2\right)
   \,\ln \, \frac{1-x}{x}
 + 8x(1-x)-1 \right] \hspace*{10mm}
\eea
in the Altarelli-Parisi splitting function $P_{q\leftarrow\gamma}^{\dis} =
P_{q\leftarrow \gamma}^{\ms} - e_q^2\, P_{q\leftarrow q}\otimes C_{\gamma}$
\cite{grv92}.

\newpage

\section{Inclusive and Diffractive Jet Production}

While at LO hadronic jets are directly identified as final-state partons,
their definition becomes subtle at higher orders, when several partons (or
hadrons) can be combined to form a jet. According to the standardization of
the 1990 Snowmass meeting, particles $i$ are added to a jet cone $J$ with
radius $R$, if they have a distance $R_i = \sqrt{(\eta_i-\eta_J)^2+(\phi_i-\phi_J)^2} < R$
from the cone center. However, these broad combined jets are difficult to
find experimentally, so that several modifications (mid-points, additional
seeds, iterations) have been successively applied by the various
experiments. The deficiencies of the cone algorithm are remedied in the
longitudinally invariant $k_T$ clustering algorithm, where one uses only the
combination criterion $R_{ij} < 1$ for any pair of particles $i$ and $j$.
Unfortunately, this algorithm scales numerically with the cubic power of the
number $N$ of the particles involved. Only recently a faster version has
been developed making use of geometrical arguments and diagrammatic methods
known from computational science \cite{cac06}. The publicly available {\tt
FastJet} code scales only with $N\ln N$ and is now rapidly adopted, in
particular for the LHC, where the particle multiplicity is high.

\begin{figure}
 \centering
 \includegraphics[width=.5\columnwidth]{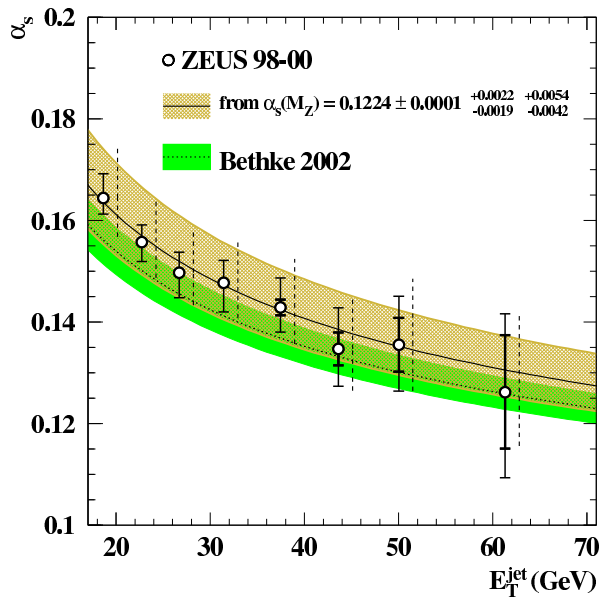}
 \caption{\label{fig:2}Strong coupling constant as measured from scaling
 violations in inclusive single-jet production at ZEUS \cite{alphas}.}
\end{figure}

Single (inclusive) jets benefit from high statistics and the presence of
a single (transverse) energy scale $E_T$, which makes them easily accessible
experimentally and predictions for them theoretically stable. The
$E_T$-distribution of the single-jet cross section can then be used to
determine e.g.\ the strong coupling constant from scaling violations, as
shown in Fig.\ \ref{fig:2}.
%
%
However, the single-jet cross section
\bea
 \hspace*{-10mm}
  \frac{\mbox{d}^2\sigma}{\mbox{d}E_T\mbox{d}\eta}
  &=& \sum_{a,b} \int_{x_{a,\min}}^1 \mbox{d}x_a x_a 
  f_{a/A}(x_a,M_a^2) x_b f_{b/B}(x_b,M_b^2)
  \frac{4E_AE_T}{2x_aE_A-E_Te^{\eta}}
  \frac{\mbox{d}\sigma}{\mbox{d}t}
 \hspace*{10mm}
\eea
includes a convolution over one of the longitudinal momentum fractions
of the partons, so that parton densities can not be uniquely determined.

%
\begin{figure}
 \centering
 \includegraphics[width=0.9\columnwidth]{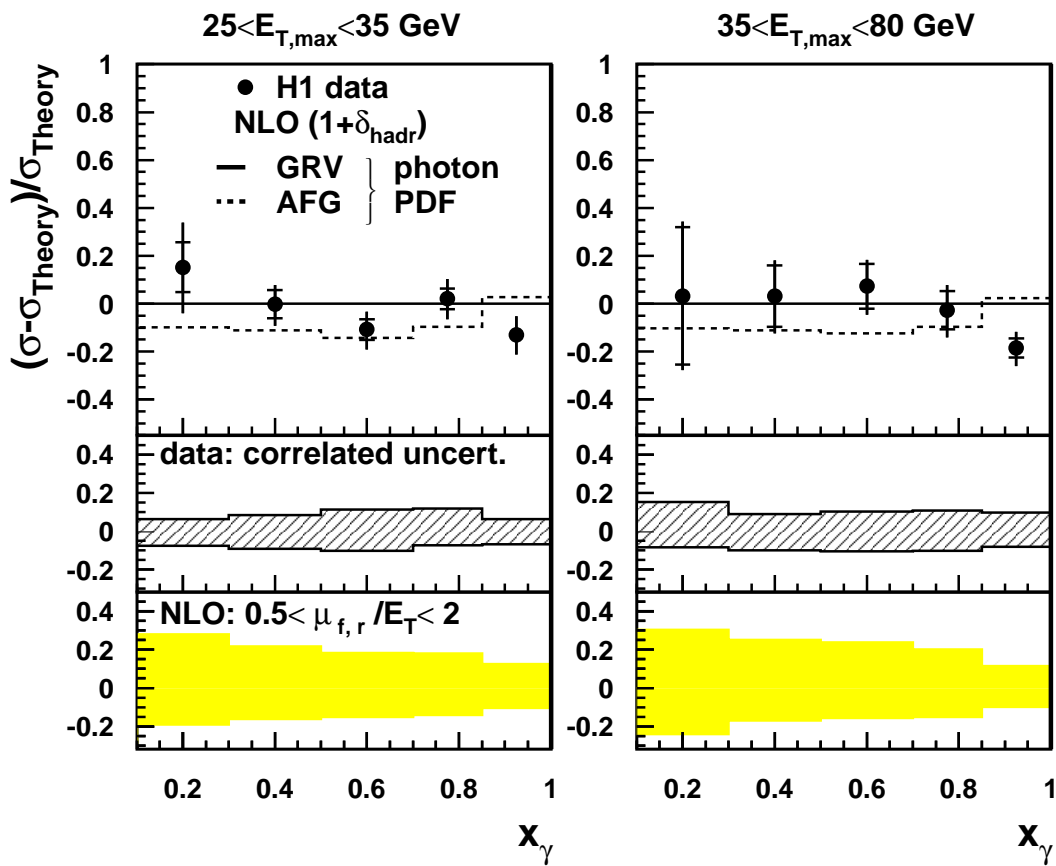}
 \caption{\label{fig:3}Sensitivity of the dijet photoproduction cross
 section as measured by H1 on the GRV and AFG parameterizations of the
 parton densities in the photon \cite{dijets}.}
\end{figure}
%
In addition to the transverse energy $E_T$ and pseudorapidity $\eta_1$ of
the first jet, the inclusive dijet cross section
\beq
  \frac{\mbox{d}^3\sigma}{\mbox{d}E_T^2\mbox{d}\eta_1\mbox{d}\eta_2}
  = \sum_{a,b} x_a f_{a/A}(x_a,M_a^2) x_b f_{b/B}(x_b,M_b^2)
  \frac{\mbox{d}\sigma}{\mbox{d}t}
\eeq
depends on the pseudorapidity of the second jet $\eta_2$. In LO only two
jets with equal transverse energies can be produced, and the observed
momentum fractions of the partons in the initial electrons or hadrons
$x_{a,b}^{\rm obs} = \sum_{i=1}^{2} E_{T_i}e^{\pm\eta_i} / (2E_{A,B})$
equal the true momentum fractions $x_{a,b}$. If the energy transfer
$y=E_\gamma/E_e$ is known, momentum fractions for the partons in photons
$x_\gamma^{\rm obs}=x_{a,b}^{ \rm obs}/y$ can be deduced. In NLO, where a
third jet can be present, the observed momentum fractions are defined by the
sums over the two jets with highest $E_T$, and they match the true momentum
fractions only approximately. Furthermore, the transverse energies of the
two hardest jets need no longer be equal to each other. Even worse, for
equal $E_T$ cuts and maximal azimuthal distance $\Delta\phi=\phi_1-\phi_2=
\pi$ the NLO prediction becomes sensitive to the method chosen for the
integration of soft and collinear singularities. The theoretical cross
section is then
strongly scale dependent and thus unreliable. This sensitivity also
propagates into the region of
large observed momentum fractions. It is thus preferable to cut on the
average $\bar{E}_T=(E_{T_1}+E_{T_2})/2$. The sensitivity of the dijet
photoproduction cross section as measured by H1 on the GRV and AFG
parameterizations of the  parton densities in the photon is shown in
Fig.\ \ref{fig:3}.

%
\begin{figure}
 \centering
 \includegraphics[width=0.66\columnwidth]{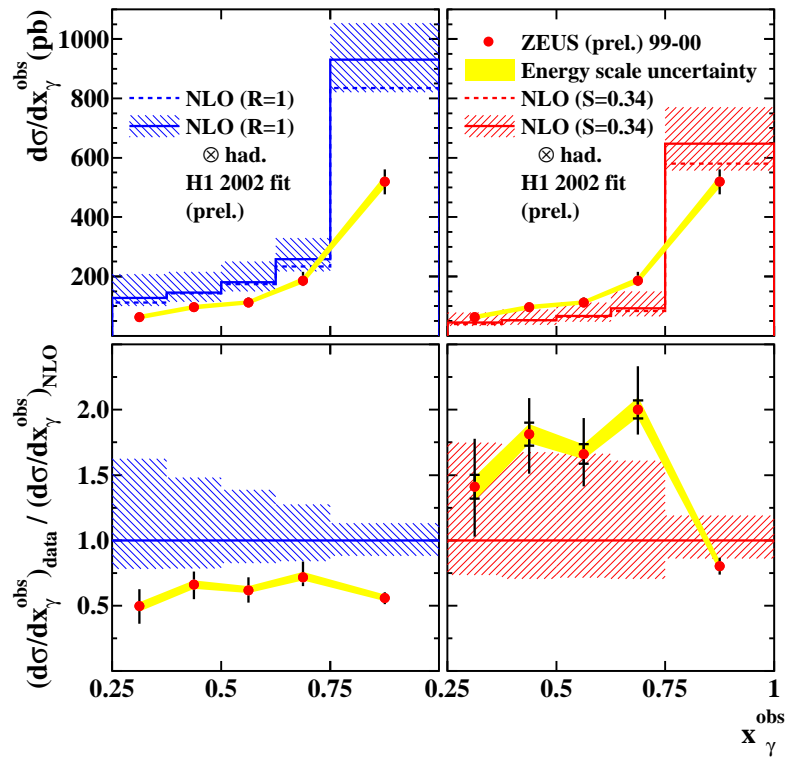}
 \caption{\label{fig:4}Dependence of the diffractive dijet cross section on
 the observed longitudinal momentum fraction of the scattered photon at
 ZEUS \cite{diffr}.}
\end{figure}
%
In diffractive processes with a large rapidity gap between a leading proton
\cite{kla04a}, neutron \cite{kla06} or some other low-mass hadronic state
and a hard central system, QCD factorization is expected to hold for
deep-inelastic scattering, so that diffractive parton densities can be
extracted from experiment, but
to break down for hadron-hadron scattering, where initial-state rescattering
can occur. In photoproduction, these two scenarios correspond to direct and
resolved processes, which are however closely related as noted in Section 1.
It is thus interesting to investigate the breakdown of factorization in
kinematic regimes where direct or resolved processes dominate. This can
either be done by measuring the dependence on the photon virtuality $Q^2$
(transition from virtual to real photons) \cite{kla04b}, $E_T$ (direct
processes are harder than resolved photons), or $x_\gamma^{\rm obs}$ (=1 for
direct processes at LO). The latter distribution is confronted in Fig.\
\ref{fig:4} with the hypothesis of no (or global) factorization breaking
(left) and with a suppression factor $S$ of 0.34 \cite{kai03} applied to 
resolved processes only (right). Note that the inter-dependence of direct
and resolved processes requires the definition of a new factorization scheme
with suppression of the scale-dependent logarithm also in the direct
contribution \cite{kla05}
\bea
 M(Q^2,S)_{\overline{\rm MS}} &=& \le-\frac{1}{2N_c} P_{q_i\leftarrow
 \gamma}(z)\ln\lr\frac{M_{\gamma}^2 z}{p_T^{*2}(1-z)}\rr+{Q_i^2\over2} \re S
 \nonumber \\
 && \ -\frac{1}{2N_c} P_{q_i\leftarrow\gamma}(z)
 \ln\lr\frac{p_T^{*2}}{zQ^2+y_s s}\rr.
\eea

\section{Light and Heavy Hadron Production}

%
\begin{figure}
 \centering
 \includegraphics[width=\columnwidth]{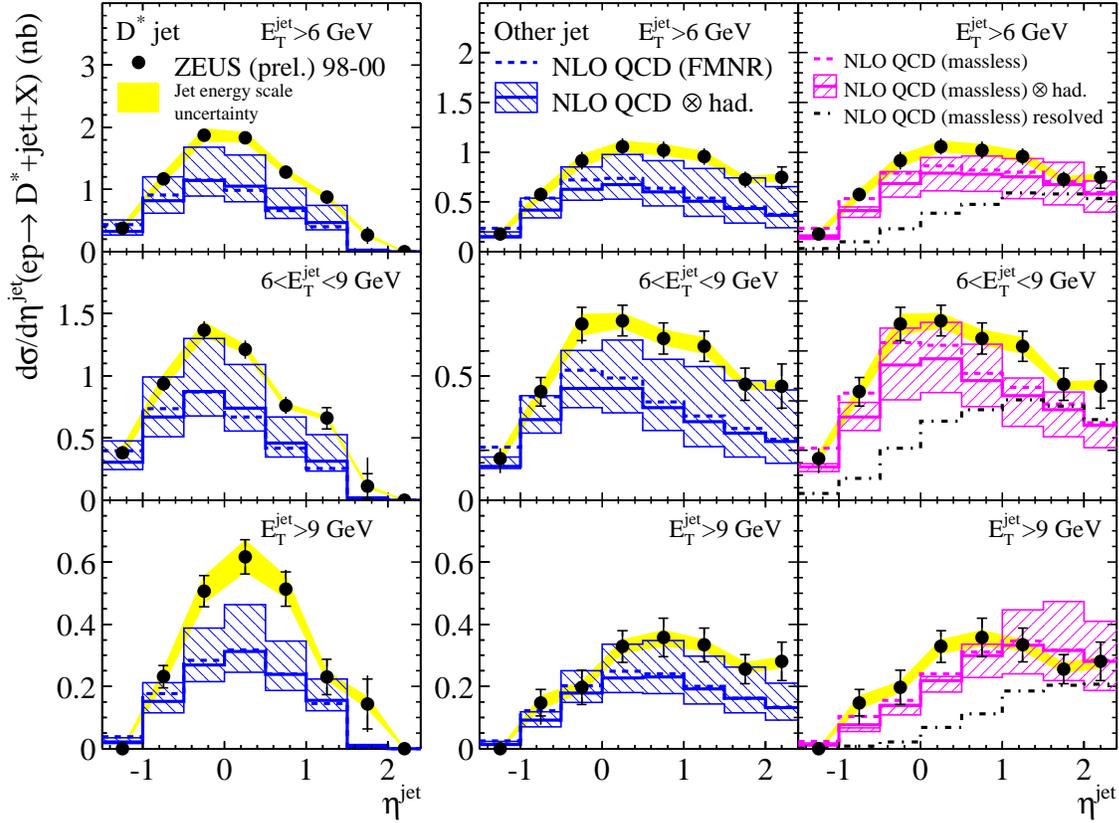}
 \caption{\label{fig:5}Rapidity distributions of $D^*$-mesons and associated
 jets as measured by ZEUS and compared to massive (fixed-order) and massless
 (variable flavor number scheme) calculations \cite{charm}.}
\end{figure}
%
If individual hadrons are experimentally identified, the cross sections
above have to be modified to include convolutions over fragmentation
functions $D(z)$. For light quarks and gluons, these non-perturbative, but
universal distributions must be fitted to $e^+e^-$ data, but then produce
successful predictions for HERA data at NLO. For heavy quarks, the
fragmentation functions can in principle be calculated perturbatively, if
the heavy-quark mass is kept finite (``fixed-order scheme''), although they
must for large $E_T$ be evolved using renormalization group equations
(for example at ``next-to-leading logarithm'') \cite{cac01}. An alternative
method is to fit the fragmentation functions for $D$- and $B$-mesons again
to $e^+e^-$ data at large $E_T$ (``variable flavor number scheme''). If in
addition the finite mass terms are kept in the hard coefficient functions,
one moves from a ``zero-mass scheme'' to a ``general-mass scheme'' and can
achieve a smooth transition from large to small $E_T$ \cite{kni05}. A
comparison of both theoretical approaches to recent $D^*$+jet data from ZEUS
is shown in Fig.\ \ref{fig:5}. While the massive calculation with central
scale choice clearly underestimates the data, the variable flavor number
scheme allows not only for direct, but also for resolved-photon
contributions and tends to give a better description of the data over the
full rapidity range. Note that both predictions have been multiplied by
hadronization corrections modeled with Monte Carlo simulations. While
several calculations for inclusive single-hadron production with real
photons are available, a theoretical investigation of the transition region
to virtual photons and of the production of two hadrons, for example in the
forward region, is still needed.

%
\begin{figure}
 \centering
 \includegraphics[width=0.66\columnwidth]{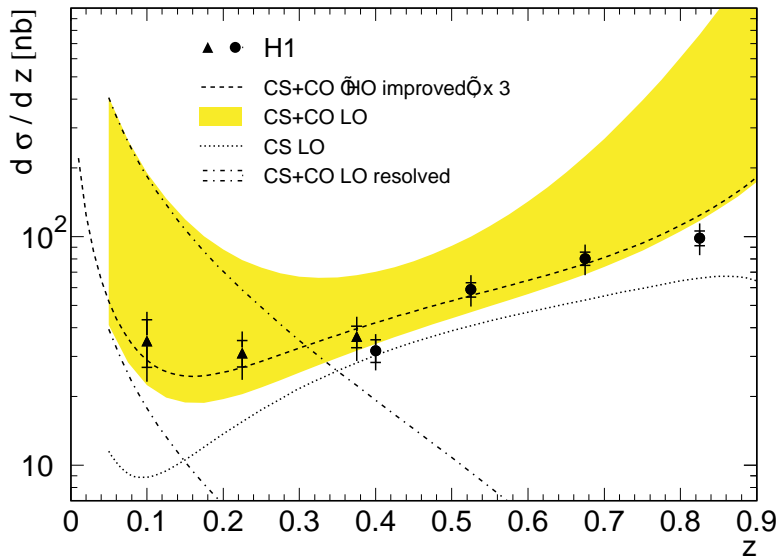}
 \caption{\label{fig:6}Direct and resolved contributions to the
 color-singlet and color-octet $J/\psi$ energy distribution in
 photoproduction at HERA \cite{jpsi}.}
\end{figure}
%
The production of heavy quark-antiquark bound states is still far from
being understood theoretically. While color-singlet (CS) states are to some
extent formed already during hard collisions, their contribution has been
shown to be both theoretically incomplete due to uncanceled infrared
singularities as well as phenomenologically insufficient due to an
order-of-magnitude discrepancy with the measured $p_T$-spectrum of
$J/\psi$-mesons at the Tevatron. On the other hand, non-relativistic QCD
(NRQCD) allows for a systematic expansion of the QCD Lagrangian in the
relative quark-antiquark velocity and for additional color-octet (CO)
contributions with subsequent color neutralization through soft gluons.
$J/\psi$-production in photon-photon collisions at LEP can then be
consistently described \cite{kla02b}, as can be the photoproduction data from
HERA in Fig.\ \ref{fig:6}. At HERA, the color-octet contribution becomes
important only at small momentum-transfer $z$ of the photon to the
$J/\psi$-meson. Unfortunately, recent CDF data do not support the prediction
of transverse polarization of the produced $J/\psi$-mesons at large $p_T$ as
predicted from the on-shell fragmentation of final-state gluons within
NRQCD, so that further experimental and theoretical studies are urgently
needed.

\section{Prompt Photon Production}

%
\begin{figure}
 \centering
 \includegraphics[width=0.66\columnwidth]{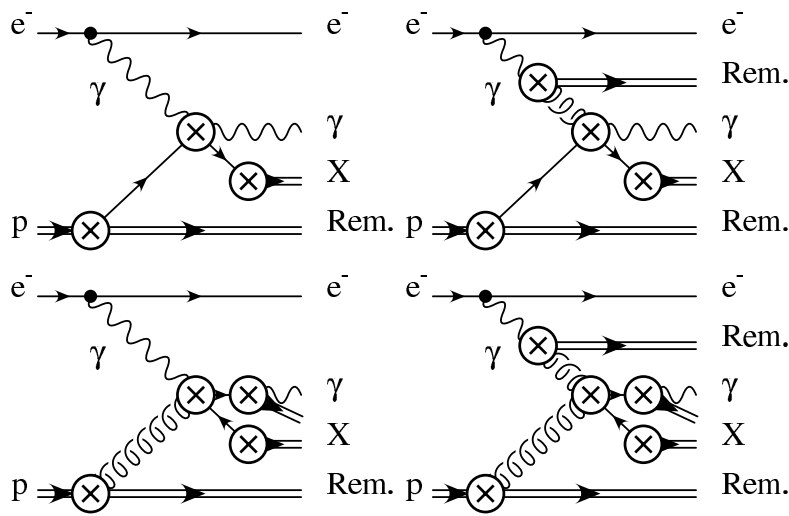}
 \caption{\label{fig:8}Factorization of photoproduction of prompt photons.}
\end{figure}
\begin{figure}
 \begin{center}
 \includegraphics[width=0.33\columnwidth,clip=]{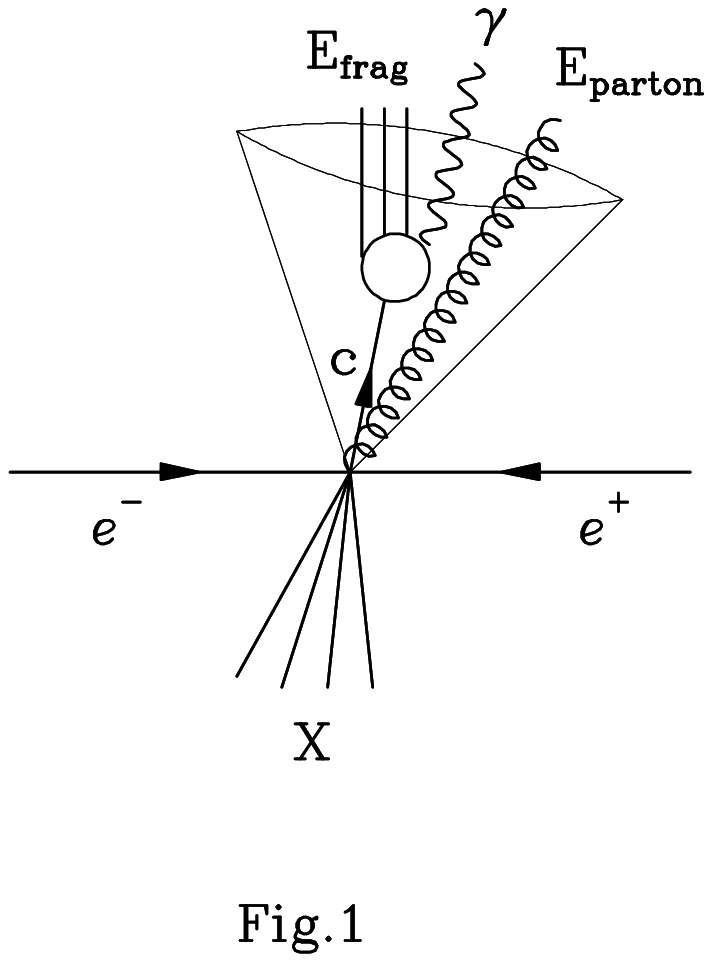}
 \caption{\label{fig:7}
 Illustration of an isolation cone containing a parton $c$ that fragments
 into a photon $\gamma$ plus hadronic energy $E_{\rm frag}$. In addition,
 a gluon enters the cone and fragments giving hadronic energy
 $E_{\rm parton}$.}
 \end{center}
\end{figure}

%
The production of prompt photons in association with jets receives
contributions from direct and resolved initial photons as well as direct
and fragmentation contributions in the final state as shown in Fig.\
\ref{fig:8}.
Photons produced via fragmentation usually lie inside hadronic jets, while
directly produced photons tend to be isolated from the final state hadrons.
The theoretical uncertainty coming from the non-perturbative fragmentation
function can therefore be reduced if the photon is isolated in phase space.
At the same time the experimental uncertainty coming from photonic decays of
$\pi^0$, $\eta$, and $\omega$ mesons is considerably reduced.
Photon isolation can be achieved by limiting the (transverse)
hadronic energy $E_{(T)}^{\rm had}$ inside a cone of size $R$ around the
photon to
\beq
 E_{(T)}^{\rm had}<\epsilon_{(T)} E_{(T),\gamma}.
\eeq
This is illustrated in Fig.\ \ref{fig:7}.
%
%
Recently an improved photon isolation criterion
\beq
 \sum_i E^{\rm had}_{(T),i} \theta(\delta-R_{i}) < \epsilon E_{(T),\gamma}
 \lr{1-\cos\delta\over 1-\cos\delta_0}\rr,
\eeq
has been proposed, where $\delta\leq\delta_0$ and $\delta_0$ is now the
isolation cone \cite{isolation}. This procedure allows the fragmentation
contribution to vanish in an infrared safe way.

%
\begin{figure}
 \centering
 \includegraphics[width=0.66\columnwidth]{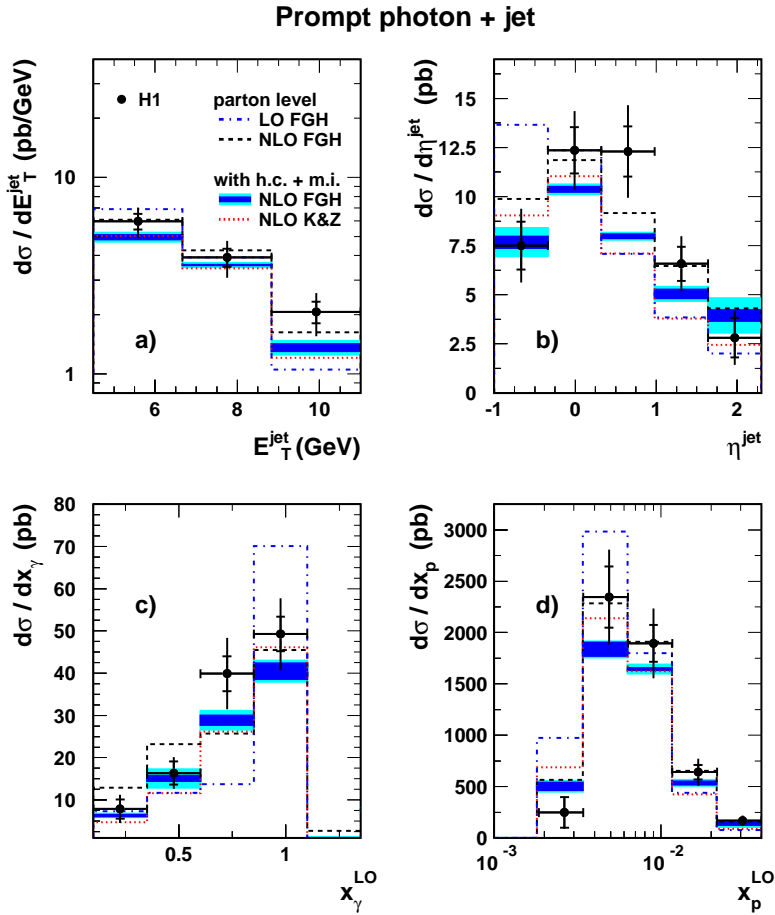}
 \caption{\label{fig:9}Various distributions of photoproduction of prompt
 photons in association with jets as measured by H1 and compared to two
 different QCD calculations \cite{phojet}.}
\end{figure}
%
Photoproduction of prompt photons and jets has been measured
by the H1 collaboration and compared with two QCD predictions, which differ
in their inclusion of NLO corrections to the resolved and fragmentation
contributions. Only after modeling hadronization corrections and multiple
interactions with Monte Carlo generators, the measured distributions shown
in Fig.\ \ref{fig:9} agree with the QCD predictions, showing the particular
sensitivity of photon final states to hadronic uncertainties.

\section{Summary}

Photoproduction processes have been abundantly measured at HERA and
stimulated many different theoretical studies, ranging from the
investigation of the foundations of QCD as inscribed in its factorization
theorems, over the determination of its fundamental parameter, the strong
coupling constant, to improvements in our understanding of proton and photon
structure as well as light and heavy hadron formation.

With the shutdown of HERA on July 1, 2007, many questions will remain
unanswered, in particular in the diffractive and non-relativistic kinematic
regimes, and this for quite some time until the eventual construction of
a new electron-hadron collider such as eRHIC or an International Linear
Collider. Photon-induced processes in ultraperipheral heavy-ion collisions
may offer a chance to continue investigations in this interesting
field, opening in addition a window to nuclear structure, if these processes
can be experimentally isolated.

\ack

The author thanks the organizers of the Trento 2007 workshop for the kind
invitation and for the creation a particularly stimulating atmosphere.




\section*{References}
\begin{thebibliography}{10}
\bibitem{kla02}For an extensive theoretical review of hard photoproduction,
 see e.g.\ Klasen M 2002 {\it Rev. Mod. Phys.} {\bf 74} 1221
\bibitem{kes75}Kessler P 1975 {\it Acta Phys. Austr.} {\bf 41} 141
\bibitem{alb02}Albino S, Klasen M and S\"oldner-Rembold S 2002 {\it Phys.
 Rev. Lett.} {\bf 89} 122004
\bibitem{grv92}Gl\"uck M, Reya E and Vogt A 1992 {\it Phys. Rev.} D {\bf 45}
 3986
\bibitem{alphas}Chekanov S et al. [ZEUS Collaboration] 2003 {\it Phys.
 Lett.} B {\bf 560} 7
\bibitem{cac06}Cacciari M and Salam G 2006 {\it Phys. Lett.} B {\bf 641} 57
\bibitem{dijets}Adloff C et al. [H1 Collaboration] 2002 {\it Eur. Phys. J} C
 {\bf 25} 13
\bibitem{diffr}Chekanov S et al. [ZEUS Collaboration], Abstract 6-249,
 submitted to ICHEP 2004, Beijing, China
\bibitem{kla04a}Klasen M and Kramer G 2004 {\it Eur. Phys. J.} C {\bf 38} 93
\bibitem{kla06}Klasen M and Kramer G 2006 {\it Preprint} hep-ph/0608235
\bibitem{kla04b}Klasen M and Kramer G 2004 {\it Phys. Rev. Lett.} {\bf 93}
 232002
\bibitem{kai03}Kaidalov A, Khoze V, Martin A and Ryskin M 2003 {\it Phys.
 Lett.} B {\bf 567} 61
\bibitem{kla05}Klasen M and Kramer G 2005 {\it J. Phys.} G {\bf 31} 1391
\bibitem{charm}Chekanov S et al. [ZEUS Collaboration], Abstract 5-332,
 submitted to ICHEP 2004, Beijing, China
\bibitem{cac01}Cacciari M, Frixione S and Nason P (2001) {\it JHEP} {\bf
 0103} 6
\bibitem{kni05}Kniehl B, Kramer G, Schienbein I and Spiesberger H 2005 {\it
 Phys. Rev.} D {\bf 71} 014018
\bibitem{jpsi}Adloff C et al. [H1 Collaboration] 2002 {\it Eur. Phys. J} C
 {\bf 25} 25
\bibitem{kla02b}Klasen M, Kniehl B, Mihaila L and Steinhauser M 2002 {\it
 Phys. Rev. Lett.} {\bf 89} 032001
\bibitem{isolation}Frixione S 1998 {\it Phys. Lett.} B {\bf 429} 369
\bibitem{phojet}Aktas A et al. [H1 Collaboration] 2005 {\it Eur. Phys. J} C
 {\bf 38} 437
\end{thebibliography}
\end{document}